\documentclass[aps,prd,amssymb,nofootinbib,floatfix,a4paper,twocolumn]{revtex4}

\usepackage{mathrsfs}
\usepackage{latexsym,amsmath,amsfonts,amssymb}
\usepackage{bm}
\usepackage{graphicx}

\allowdisplaybreaks

\newcommand{\be}{\begin{equation}}
\newcommand{\ee}{\end{equation}}
\newcommand{\e}{\widehat{\mathcal E}_{\rm eff}}
\newcommand{\hHf}{{{\widehat H}_{\rm eff}}}

\begin{document}

\title{Novel approach to binary dynamics: application to the fifth post-Newtonian level}

\author{Donato Bini$^{1,2}$, Thibault Damour$^3$, Andrea Geralico$^1$}
  \affiliation{
$^1$Istituto per le Applicazioni del Calcolo ``M. Picone,'' CNR, I-00185 Rome, Italy\\
$^2$INFN, Sezione di Roma Tre, I-00146 Rome, Italy\\
$^3$Institut des Hautes \'Etudes Scientifiques, 91440 Bures-sur-Yvette , France.
}

\date{\today}

\begin{abstract}
We introduce a new methodology for deriving the conservative dynamics of gravitationally interacting binary systems. Our approach
combines, in a novel way, several theoretical formalisms: post-Newtonian, post-Minkowskian, multipolar-post-Minkowskian, gravitational self-force,
and effective one-body. We apply our method to the derivation of the fifth post-Newtonian dynamics. By restricting
our results to the third post-Minkowskian level, we give the first independent confirmation of the recent result of  Bern {\it et al.} 
 [Phys.\ Rev.\ Lett.\  {\bf 122}, 201603 (2019)]. We also offer checks for future  fourth post-Minkowskian calculations.
 Our technique can, in principle, be extended to higher orders of perturbation theory.
\end{abstract}

\maketitle

{\it Introduction.}---The detection of  the coalescence of compact binaries by the LIGO-Virgo collaboration~\cite{LIGOScientific:2018mvr}
demands an ever more accurate knowledge of the gravitational dynamics and radiation
of binary systems. We propose here a new methodology for improving the theoretical description of the conservative dynamics
of two-body systems in General Relativity. This methodology unifies in a novel way an array of previously
developed theoretical tools, and combines it with some new insights. 
This allows one to reach in an expedient manner new high-order
results of direct physical significance. Here, we exemplify the efficiency of our method by
applying it to  the first (essentially complete) computation of  the conservative 
two-body dynamics (of two non-spinning masses $m_1, m_2$)
at the {\it fifth post-Newtonian} (5PN) accuracy, {\it i.e.} one order in $(v/c)^2$ beyond the last 
post-Newtonian (PN) order at which this dynamics has been heretofore fully derived~\cite{Damour:2014jta,Jaranowski:2015lha,Bernard:2015njp,Damour:2016abl,Marchand:2017pir,Foffa:2019rdf,Foffa:2019yfl}. [Our 5PN-level result cannot be compared with
the recent 5PN-level works~\cite{Foffa:2019hrb,Blumlein:2019zku}, because the latter have  computed only the  small,
and non gauge-invariant, subset of ``static''  contributions to the 5PN Hamiltonian.]  As a by product of our calculation, we also compute
the 5PN-level contribution to the (gauge-invariant) scattering angle of two bodies considered at the {\it third post-Minkowskian}
(3PM) approximation. We find a result which is in agreement with the corresponding result recently derived from a
two-loop scattering amplitude computation~\cite{Bern:2019nnu,Bern:2019crd}, thereby providing the first 
confirmation of the latter result going beyond the 4PN-level checks derived from the gauge-invariant 4PN scattering~\cite{Bini:2017wfr}.
The result presented here is only a first application of a general methodology which can be extended to higher PN orders.

Let us motivate our new approach  by considering the state of the art of 
the general relativistic two-body problem.
The PN formalism has been the method of choice, during many years, for analytically tackling the dynamics of binary systems.
However, it has recently reached a level of complexity which renders further progress acutely difficult. Most of the
technical efficiency of the PN formalism comes from the fact that it systematically replaces the four-dimensional relativistic 
propagator 
\be
P_4(t, {\bf x}, t', {\bf x'}) \equiv \Box^{-1}= (\Delta - c^{-2} \partial_t^2)^{-1}\,,
\ee
entering the post-Minkowskian (PM) formalism,  
 by its formal expansion in inverse powers of the velocity of light:
\be 
\label{PNP4}
P_4^{\rm PN}(t, {\bf x}, t', {\bf x'}) =  \left(\Delta^{-1} + \frac{1}{c^2} \partial_t^2  \Delta^{-2}+ \cdots\right) \delta(t-t')\,.
\ee
Here, we consider the time-symmetric propagator (as appropriate to the derivation of the conservative dynamics). Each term
on the right-hand side (RHS) of the PN expansion \eqref{PNP4} of $P_4(t, {\bf x}, t', {\bf x'})$ is local-in-time, in the sense that it involves
a derivative of $ \delta(t-t')$. Inserting the PN expansion \eqref{PNP4} in the reduced action describing (after having integrated out the gravitational field) the PM-expanded dynamics of two
worldlines~\cite{Damour:1995kt} allows one to express the relativistic gravitational
interaction of two particles in terms of iterated integrals (given by Feynman-like diagrams) involving only the concatenation of 
the instantaneous three-dimensional propagator
\be
P_3(t, {\bf x}, t', {\bf x'})= \delta(t-t') \Delta^{-1}= - \frac{1}{4 \pi} \frac{\delta(t-t')}{| {\bf x}- {\bf x'} |}\,,
\ee 
or of its descendants $\frac{1}{c^2} \partial_t^2  \Delta^{-2} \delta(t-t') + \cdots$.
The use of such a PN-expanded propagator, together with the corresponding PN simplification of the nonlinear
vertices generated by the Einstein-Hilbert action, leads to drastic simplifications (compared to a corresponding PM-expanded action involving the original
4-dimensional propagator $P_4^{\rm PN}(t, {\bf x}, t', {\bf x'})$) in the computation  of the reduced
action, especially when using the Arnowitt-Deser-Misner Hamiltonian approach~\cite{Schafer:2018kuf}.
Indeed, the concatenated massless propagators $ \Delta^{-n}$  lead to
3-dimensional integrals containing only one length scale, namely (in ${\bf x}$-space) the distance 
$r_{12} \equiv | {\bf x}_1- {\bf x}_2 |$ between the two bodies.

However, as had been anticipated years ago~\cite{Blanchet:1987wq}, the PN approach undergoes a fundamental 
conceptual (and technical) breakdown at the fourth post-Newtonian (4PN) level. At this level the naive
PN expansion \eqref{PNP4} of the 4-dimensional propagator $P_4^{\rm PN}(t, {\bf x}, t', {\bf x'})$ 
 becomes fundamentally inadequate because the gravitational interaction necessarily involves {\it nonlocal-in-time} effects
that cannot be described in terms of the sequence of quasi-instantaneous terms appearing on the RHS of  \eqref{PNP4}.
This conceptual failure of the PN expansion leads, at the technical level, to the appearance of infrared  
logarithmic divergences in the formal computation of the PN-expanded action. The current direct perturbative
computations (using the expansion \eqref{PNP4}) of the 4PN-level reduced action~\cite{Damour:2014jta,Jaranowski:2015lha,Bernard:2015njp,Damour:2016abl,Marchand:2017pir,Foffa:2019rdf,Foffa:2019yfl} have succeeded in solving this
problem in various ways.
However, this variety of approaches, which included discrepant intermediate results~\cite{Bernard:2015njp} before complete agreement was reached, show that straightforward perturbative PN computations have reached their limit of 
easily verifiable reliability. This makes it urgent to develop a new methodology, as we do here. 
Our new approach consists of several steps that we explain in turn.

{\it First step: computing the nonlocal-in-time piece of the action}---
The first step is to use results derived within the (PN-matched~\cite{Blanchet:1987wq,Blanchet:1989ki,Poujade:2001ie}) 
multipolar-post-Minkowskian (MPM) formalism~\cite{Blanchet:1985sp} to decompose, at some given PN accuracy,
the complete,  (reduced) two-body conservative action $S_{\rm tot}$  
in two separate pieces: a nonlocal-in-time part, say $S^{\rm nonloc}$, 
and a local-in-time part,  $S^{\rm loc}$:
\begin{eqnarray} 
\label{Sdecomp}
S_{\rm tot}^{\leq n \rm PN}[x_1(s_1), x_2(s_2)]&=& S_{\rm loc}^{\leq n \rm PN}[x_1(s_1), x_2(s_2)]\nonumber\\
&+ & S_{\rm nonloc}^{\leq n \rm PN}[x_1(s_1), x_2(s_2)]\,.
\end{eqnarray} 
Here, $S_{\rm tot}[x_1(s_1), x_2(s_2)]$ is a time-symmetric functional of the two worldlines  defined (before considering it approximate estimation at the $n$PN accuracy) by its PM-expansion~\cite{Damour:1995kt}. The decomposition \eqref{Sdecomp} makes sense, at a given PN accuracy, because the MPM formalism yields an efficient tool for
computing the non-local piece $ S_{\rm nonloc}^{\leq n \rm PN}$. From Ref.~\cite{Blanchet:1987wq} one knows that 
$ S_{\rm nonloc}^{\leq n \rm PN}$ starts at the 4PN level. The 4PN-accurate value of  $ S_{\rm nonloc}$ was
obtained in Ref.~\cite{Damour:2014jta} (see also the related works~\cite{Foffa:2011np,Ross:2012fc,Galley:2015kus}). The 5PN-accurate value of  $ S_{\rm nonloc}$ was obtained in section IXA of ~\cite{Damour:2015isa} (based on Ref.~\cite{Damour2010}). [See also Refs.~\cite{Blanchet:2010zd,LeTiec:2011ab} 
for the related 5PN logarithmic terms.] It reads, from Eq. (9.12) of 
~\cite{Damour:2015isa} (see also the related recent work ~\cite{Foffa:2019eeb}), 
\begin{eqnarray} 
\label{Snonloc}
S_{\rm nonloc}^{4+5 \rm PN}[x_1(s_1), x_2(s_2)]&=& \frac{G^2 {\cal M}}{c^3} \int dt {\rm Pf}_{2 r_{12}^h(t)/c}\times  \nonumber\\
&& \int  \frac{ dt'}{|t-t'|} {\cal F}_{1 \rm PN}^{\rm split}(t,t')\,.
\end{eqnarray} 
Here, ${\cal M}$ denotes the total conserved mass-energy of the binary system, while ${\cal F}_{1 \rm PN}^{\rm split}(t,t')$ denotes
the time-split version of the fractionally 1PN-accurate gravitational-wave energy flux emitted by the system, namely (using
a superscript in parenthesis to denote a repeated time-derivative)
\begin{eqnarray}
 &&{\cal F}_{1 \rm PN}^{\rm split}(t,t')= \frac{G}{c^5} \left( \frac15 I_{ab}^{\rm (3)}(t) I_{ab}^{\rm (3)}(t') \right. \nonumber\\
 &&\left.+  
  \frac1{189 c^2} I_{abc}^{\rm (4)}(t) I_{abc}^{\rm (4)}(t') +\frac{16}{45 c^2} J_{ab}^{\rm (3)}(t) J_{ab}^{\rm (3)}(t') \right).
 \end{eqnarray}
The  mass and spin multipole moments $I_{ab}$, $I_{abc}$, $J_{ab}$, entering the latter expression
are the Blanchet-Damour (1PN-accurate) source multipole moments  defined by explicit integrals over the 
stress-energy tensor of the source~\cite{Blanchet:1989ki}. Their explicit expressions for a binary system can be found
in Ref.~\cite{Blanchet:1989cu}. Eq. \eqref{Snonloc} defines an explicit functional of the two worldlines, and subtracting it from the
(in principle PM-computable) total action $S_{\rm tot}$ {\it defines} the local-in-time contribution 
$S_{\rm loc}^{\leq 5 \rm PN}$ to the two-body dynamics. The time-scale entering the
partie finie operation (Pf) used in \eqref{Snonloc} to define the logarithmically divergent integral over $t'$
 has been fixed to be $2 r_{12}^h/c$, where $r_{12}^h$ denotes
the (harmonic-coordinate) radial distance between the two bodies. Note that the meaning here of $S_{\rm loc}$ (and its corresponding $H_{\rm loc}$)  
differs from the one in Refs. \cite{Damour:2015isa,Bini:2017wfr},  where $H_{\rm loc}$ included logarithmic contributions in its definition.

{\it Second step: computing the $O(\nu)$ piece of the time-averaged redshift $\langle z_1 \rangle$
to sixth order in eccentricity}---
The second step of our approach consists in using gravitational Self-Force (SF) theory to compute to sufficient accuracy
the first-order-self-force (1SF) contribution, say $\delta z_1= O(\nu)$, to the time-averaged redshift $\langle z_1 \rangle= \langle ds_1/dt \rangle$ of the first body,
considered as a function of the symmetric mass ratio $\nu$ and of the dimensionless radial and azimuthal frequencies $M \Omega_r$, $M \Omega_\phi$
of {\it eccentric} orbits~\cite{Detweiler:2008ft,Barack:2011ed}. 
[We denote  $M \equiv m_1+m_2$, $\mu= m_1 m_2/M$, $\nu= \mu/M= m_1 m_2/(m_1+m_2)^2$.] 
Ref.~\cite{Bini:2013zaa} has developed efficient tools for analytically computing $\delta z_1$ as a function of the inverse parameter 
of the elliptical orbit, $u_p=GM/(c^2p)$, and of the eccentricity, $e$. Current results reached either high-orders in $e$ limited to 4PN accuracy~\cite{Hopper:2015icj}, or high PN accuracy limited to fourth order in $e$~\cite{Bini:2016qtx}. 
We crucially needed, for the present work, to extend the computation of $\delta z_1$ to the sixth order in $e$,
and, to, at least, the 5PN accuracy, i.e., the sixth order in $u_p$.  The result of our computation for the coefficient of $e^6$ in $\delta z_1$ reads, at 5PN accuracy, 
\begin{eqnarray} \label{dz1}
\delta z_1^{e^6}&=& \nu\left[ \frac14 u_p^3 +\left(-\frac{53}{12}-\frac{41}{128}\pi^2\right)u_p^4 \right. \nonumber\\
&&\left.  +C_5 u_p^5 + C_6 u_p^6 + o( u_p^6)\right]+ O(\nu^2)\,,
\end{eqnarray}
where 
\begin{eqnarray}
C_5&=&-\frac{38471}{360}+\frac{6455}{4096}\pi^2-\frac{178288}{5}\ln(2)\nonumber\\
&&+\frac{1994301}{160}\ln(3)+\frac{1953125}{288}\ln(5)+16\gamma+8\ln(u_p)\,, \nonumber\\ 
C_6&=&-\frac{17344111}{5040}+\frac{782899}{4096}\pi^2+\frac{66668054}{135}\ln(2)\nonumber\\
&& -\frac{29268135}{448}\ln(3)-\frac{2027890625}{12096}\ln(5)\nonumber\\
&&-\frac{1694}{5}\gamma-\frac{847}{5}\ln(u_p) \,. 
\end{eqnarray}
We have also determined the higher-order contributions in $u_p$ up to $u_p^{19/2}$ ~\cite{BDG_2019}.

{\it Third step: using the first law of binary dynamics to translate  $\delta z_1^{e^6}$ into a
corresponding $O( p_r^6)$-accurate,  effective-one-body Hamiltonian}---
The first law of binary dynamics~\cite{LeTiec:2011ab,Barausse:2011dq,Tiec:2015cxa} allows one to transform
the  gauge-invariant information contained in our new result \eqref{dz1} (together with the previous $O(e^4)$ results~\cite{Hopper:2015icj,Bini:2016qtx})
into a corresponding knowledge of the (gauge-fixed)
two-body Hamiltonian, as expressed in  effective-one-body (EOB) theory~\cite{Buonanno:1998gg,Damour:2000we}.
To do this we had to extend the results of ~\cite{Tiec:2015cxa} to the sixth order in the ($\mu$-rescaled) 
radial momentum $p_r$. EOB theory expresses the two-body Hamiltonian $H$ ($={\mathcal M} c^2$) in terms of a rescaled ``effective'' Hamiltonian 
$\hHf $ according to
\be \label{Heob}
H= Mc^2 \sqrt{ 1 + 2\nu (\hHf -1)}\,.
\ee
In turn, $\hHf$ is expressed in terms of various bricks: two radial potentials $A(u; \nu)$, and ${\bar D}(u; \nu) \equiv (A(u; \nu) B(u; \nu))^{-1}$,
and a momentum-dependent potential $Q(u,p;\nu)$, where $u\equiv GM/(c^2r)$. Namely,  henceforth setting $c=1$,
\be \label{Heff}
\hHf^2= A(u;\nu)[1+A(u;\nu) \bar D (u;\nu) p_r^2 + p_\phi^2 u^2 +Q(u,p;\nu)]\,.
\ee
The PN expansions of the potentials $A(u;\nu)$, and ${\bar D}(u;\nu)$ are written as $A(u;\nu) = 1-2u +\sum_n a_{n}(\nu, \ln u) u^n$
and  ${\bar D}(u;\nu)= 1 + \sum_n {\bar d}_n(\nu, \ln u) u^n$.
In the gauge (hereafter called ``$p_r$ gauge'') introduced in ~\cite{Damour:2000we}, the PN expansion of $Q(u,p)$ is given by a double expansion in
$u$ and $p_r^2$, say $Q=q_4(u;\nu) p_r^4+q_6(u;\nu) p_r^6+q_8(u;\nu) p_r^8 + \cdots$, where $q_m(u;\nu) = \sum_n q_{mn}(\nu, \ln u) u^n$.  In addition, all the (logarithmically running) $\nu$-dependent coefficients $a_{n}(\nu, \ln u)$, ${\bar d}_n(\nu, \ln u)$,  $q_{mn}(\nu, \ln u)$ are {\it polynomials in} $\nu$,
starting at $\nu^1$, and of degree increasing with $n$. We derived the relation linking  the 1SF ($O(\nu)$) piece in $q_6(u;\nu)$
to the 1SF redshift $\delta z_1(u_p, e)=\delta z_1^{e^0}(u_p)  + \delta z_1^{e^2}(u_p) e^2 +\delta z_1^{e^4}(u_p) e^4 +\delta z_1^{e^6}(u_p) e^6$. This allowed us to extend the previous 1SF knowledge of $q_4(u;\nu) p_r^4$~\cite{Bini:2016qtx}  to the $p_r^6$ level, namely
\be
\label{q61SF}
q_6(u;\nu) = \nu q_{62}^{\nu^1}u^2 +\nu q_{63}^{\nu^1}u^3 
+ O(u^{7/2}) + O(\nu^2)\,,
\ee
where $q_{62}^{\nu^1}$ is a known 4PN term~\cite{Damour:2015isa} and where
\begin{eqnarray}
q_{63}^{\nu^1}&=& \frac{2613083}{1050}+\frac{6875745536}{4725}\ln(2)\nonumber\\
&-&\frac{23132628}{175}\ln(3)-\frac{101687500}{189}\ln(5)\,,
\end{eqnarray}
is a new, 5PN level, result.
See Ref.~\cite{BDG_2019} for the higher-order contributions in $u$ (up to $u^{19/2}$ included).

{\it Fourth step:  determining the 1SF contribution to the {\it local-in-time} 5PN-accurate Hamiltonian by subtracting the nonlocal action}---
Inserting our new result \eqref{q61SF}, together with the previous high-PN 1SF knowledge of $A(u; \nu)$, ${\bar D}(u; \nu)$ and $q_4(u;\nu)$,
in Eqs. \eqref{Heob} and \eqref{Heff} determines the two-body Hamiltonian at the  combined 1SF $+$ 5PN accuracy. [At the level of the
unrescaled, total Hamiltonian $H$, Eq. \eqref{Heob}, 1SF accuracy means knowing both the $\nu^1$  and the $\nu^2$ contributions.]
We can then subract from the full Hamiltonian action $\int p dq - H(q,p) dt$ the nonlocal-in-time term \eqref{Snonloc} to compute
the local-in-time Hamiltonian action $\int p dq - H_{\rm loc}(q,p) dt$. This is conveniently done by using the Delaunay averaging technique
of the nonlocal action introduced in ~\cite{Damour:2015isa}. This averaging technique leads to a gauge-invariant result which can then be
expressed in the EOB-$p_r$ gauge. The so obtained 1SF $+$ 5PN accurate local-in-time Hamiltonian $H_{\rm loc}(q,p)$ can then be expressed (via the
universal EOB energy map \eqref{Heob})  in terms of corresponding 1SF $+$ 5PN accurate EOB potentials $A_{\rm loc}(u; \nu)$, 
${\bar D}_{\rm loc}(u; \nu)$ and $Q_{\rm loc}(u, p_r ;\nu)$. All logarithmic dependence (including numerical logs, like $\ln 2$) has disappeared from these local potentials. For instance, the local contribution to $q_6(u;\nu)$ was found to be 
$q_{6}^{ \rm loc}= -\frac{9}{5}\nu u^2 + \frac{123}{10}\nu u^3 + O(\nu^2)$. Here, the contribution $ +\frac{123}{10}\nu u^3$
is at the 5PN level.

{\it Fifth step:  using EOB-PM theory to determine most of the nonlinear dependence on $\nu$ of the local Hamiltonian}---
At this stage, while we know the exact dependence of the nonlocal action \eqref{Snonloc} on the two masses $m_1, m_2$,
and therefore on $\nu$ for a given $M$, our use of  SF technology has limited our determination of the local Hamiltonian  
$H_{\rm loc}(q,p; \nu)$ to the 1SF accuracy: $H_{\rm loc}= M c^2+ \nu H^{(1)}_{\rm loc} + \nu^2 H^{(2)}_{\rm loc}+ O(\nu^3)$.
We can, however, determine most of the higher-order powers in $\nu$ by using results from the EOB formalism applied 
to PM-expanded scattering. More precisely, we can use two constraints.

On the one hand, the exact $\nu$ dependence of the EOB Hamiltonian has been determined 
both at the first post-Minkowskian (1PM) level~\cite{Damour:2016gwp},
and at the second post-Minkowskian (2PM) level~\cite{Damour:2017zjx}. By transforming the latter results (obtained
in a special ``energy'' gauge) into the (standard) EOB-$p_r$ gauge used above, we can determine the exact $\nu$ dependence
of the 5PN-accurate (local and nonlocal) Hamiltonian for all the terms in the Hamiltonian which are either $\propto u^1$ or
$\propto u^2$. For instance, we thereby found that the coefficient $q_{82}$ of $p_r^8 u^2$ in the $Q$ potential is
$q_{82}=q_{82}^{\rm loc} = \frac67 \nu+\frac{18}7 \nu^2+\frac{24}{7}\nu^3-6\nu^4$.

On the other hand, the general dictionary ~\cite{Damour:2017zjx} between the EOB Hamiltonian and the PM-expanded scattering function,
$\frac12 \chi(\e,j)= \sum_n\chi_n(\e)/j^n$, where $j\equiv J/(G m_1 m_2)$, has recently been used~\cite{Damour2019}
to show that the combination $ \left(1 + 2\nu (\e -1) \right)^{n-1} \chi_n(\e)$ was a polynomial in $\nu$ of degree
$d(n)$ equal to the integer part of $(n-1)/2$. This yields a strong restriction on the $\nu$ dependence of the
coefficients of the 5PN-level local Hamiltonian, 
\be \label{Hloc5pn}
H_{\rm loc}^{5 \rm PN}=\sum_{m + n=6} h_{2m \, n}(\nu) (p^2)^m u^n \,.
\ee
[For notational simplicity, we use in Eq. \eqref{Hloc5pn} $p^2$ to denote either $p_r^2$ or $p_t^2 \equiv p_\phi^2/r^2$.]
In order to apply this restriction, we computed (as a function of the coefficients $h_{2m \, n}(\nu)$)
the scattering angle implied by the total, 5PN-accurate Hamiltonian (using the technique of Ref.~\cite{Bini:2017wfr}).

{\it Final result for the local-in-time 5PN-accurate Hamiltonian}---
The 5PN-level local Hamiltonian, Eq. \eqref{Hloc5pn}, {\it a priori} contains (in our gauge)  36 unknown numerical coefficients, say $h_{2m \, n}^{\nu^k}$ 
parametrizing the powers of $\nu$ in the various coefficients $h_{2m \, n}(\nu) = \sum_{k=1}^{k_{\rm max}(m, n)} h_{2m \,n}^{\nu^k} \nu^k$ appearing in Eq. \eqref{Hloc5pn}. [Here, we do not distinguish the coefficients of $p_r^2$ or $p_t^2$. If distinguished, there are 108 coefficients.] The degrees of these polynomials in $\nu$ are indeed found (when $m + n=6$)
to be all equal to $ k_{\rm max}= 6$ when $n= 6-m= 1, \ldots,6$.

\begin{figure}
\includegraphics[scale=0.25]{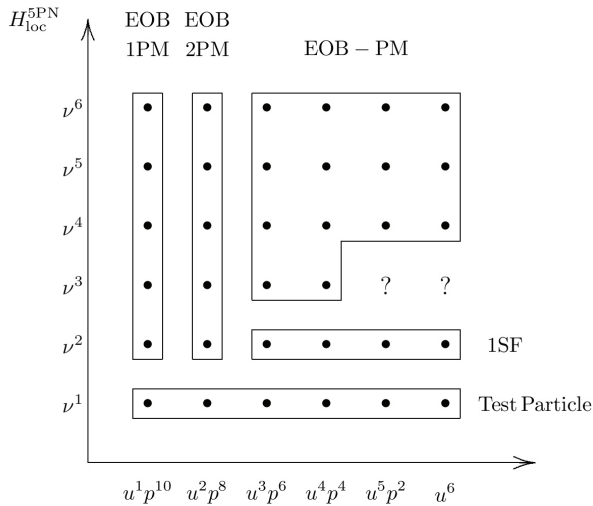}
\caption{Schematic representation of the theoretical tools used to obtain the various contributions to the 5PN-accurate local Hamiltonian. 
These contributions are keyed,  on the horizontal axis, by powers of $u=GM/r$ and squared momentum  
$p^2 \sim p_r^2 \sim p_t^2$, and, on the 
the vertical axis, by powers of $\nu \equiv m_1 m_2/(m_1+m_2)^2$. The bullets indicate the coefficients determined for the first time in the present work. The question marks denote
the only two missing coefficients.}
\end{figure}

Combining all the previous tools and results, we were able to determine 34 of the {\it a priori} unknown numerical coefficients $h_{2 m \, n}^{\nu^k}$. Fig. 1 indicates the source of information having allowed us to determine each one of these 34 coefficients:
the test-particle limit determines the  $\nu^1$ row; the 1SF computations determine the  $\nu^2$ row; the first two columns
are respectively determined by the 1PM and 2PM exact EOB Hamiltonians; the  $\nu^{\geq 3}$ dependence of the next third and fourth
columns (respectively corresponding to 3PM and 4PM) are completely determined by the EOB-PM scattering constraint mentioned
above. The latter constraint determines the coefficients in the last two columns (5PM and 6PM) except for the two coefficients
$h_{2 \,5}^{\nu^3}$ and $h_{0 \,6}^{\nu^3}$. [Distinguishing $p_r^2$ and $p_t^2$, we determine 106 coefficients among 108.] When using the (more compact) EOB parametrization of the local Hamiltonian
the full description of the local-in-time 5PN-accurate Hamiltonian $H_{\rm loc}^{5 \rm PN}$ is obtained 
by inserting in the EOB map \eqref{Heob} the  effective Hamiltonian $\hHf$ defined
by the following (logarithm-free) values of
the local pieces of the EOB building blocks $A(u; \nu)$,  ${\bar D}(u; \nu)$ and $Q(u,p_r;\nu)$:
\begin{eqnarray}
A_{\rm loc} &=& 1-2 u+2\nu u^3+\nu \left(\frac{94}{3}-\frac{41}{32}\pi^2\right)u^4\nonumber\\
&&+a_5^{\rm loc} u^5+a_6^{\rm loc} u^6\,,\nonumber\\ 
\bar D_{\rm loc} &=& 1+6\nu u^2+(52\nu-6\nu^2) u^3+\bar d_4^{\rm loc} u^4+\bar d_5^{\rm loc} u^5\,,\nonumber\\
Q_{\rm loc}&=& p_r^4 [2  (4-3\nu)\nu u^2+q_{43}^{\rm loc}u^3 +q_{44}^{\rm loc}u^4 ]\nonumber\\
&&+p_r^6 (q_{62}^{\rm loc}u^2 +q_{63}^{\rm loc}u^3 )+q_{82}^{\rm loc}p_r^8  u^2 \,,
\end{eqnarray}
with
\begin{eqnarray}
a_5^{\rm loc}&=&\left(-\frac{4237}{60}+\frac{2275}{512} \pi^2\right)\nu +\left(\frac{41}{32}\pi^2-\frac{221}{6}\right)\nu^2\,,\nonumber\\
a_6^{\rm loc}&=& \left(-\frac{1026301}{1575}+\frac{246367}{3072}\pi^2 \right)\nu +a_6^{\nu^2}\nu^2\nonumber\\
&& +4\nu^3\,,
\end{eqnarray}
\begin{eqnarray}
\bar d_4^{\rm loc}&=& \left(\frac{1679}{9}-\frac{23761}{1536}\pi^2\right)\nu +\left(-260+\frac{123}{16}\pi^2\right)\nu^2\,,\nonumber\\
\bar d_5^{\rm loc}&=&\left(\frac{331054}{175} -\frac{63707}{512}\pi^2\right) \nu+\bar d_5^{\nu^2}\nu^2\nonumber\\
&+& \left(\frac{1069}{3}-\frac{205}{16}\pi^2 \right)\nu^3\,,
\end{eqnarray}
and
\begin{eqnarray}
q_{43}^{\rm loc} &=& 20\nu-83\nu^2+10\nu^3\,,\nonumber\\ 
q_{44}^{\rm loc} &=& \left(\frac{1580641}{3150}-\frac{93031}{1536}\pi^2\right) \nu \nonumber\\
&+& \left(-\frac{2075}{3}+\frac{31633}{512}\pi^2\right)\nu^2+\left(640-\frac{615}{32}\pi^2\right)\nu^3\,,\nonumber\\ 
q_{62}^{\rm loc} &=& -\frac{9}{5}\nu-\frac{27}{5}\nu^2+6\nu^3\,,\nonumber\\ 
q_{63}^{\rm loc} &=& \frac{123}{10}\nu-\frac{69}{5}\nu^2+116\nu^3-14\nu^4\,,\nonumber\\
q_{82}^{\rm loc} &=& \frac{6}{7}\nu+\frac{18}{7}\nu^2+\frac{24}{7}\nu^3-6\nu^4\,.
\end{eqnarray}
Modulo the two undetermined coefficients  $a_6^{\nu^2}$ and $\bar d_5^{\nu^2}$,
the full 5PN-accurate dynamics is given by adding to the local action defined by $H_{\rm loc}^{\leq 5 \rm PN}$ the 4+5PN nonlocal one 
Eq. \eqref{Snonloc}.

{\it New results at 3PM and 4PM}---
As one can see on Fig. 1, our results give a {\it complete} description of the 5PN dynamics at the 3PM and 4PM levels
(fourth and fifth columns in Fig. 1). This means in particular that our findings allow us to compute, with 5PN accuracy, the 3PM
($O(G^3)$) and 4PM ($O(G^4)$) terms, $\chi_3$ and $\chi_4$, in the scattering angle. The computation at 5PN accuracy of 
$\chi_3$ from our results for the full loc + nonloc dynamics (with $\chi_3^{\rm nonloc}=0$ ~\cite{Bini:2017wfr}) yields (denoting $p_\infty\equiv \sqrt{\e^2-1}$)
\begin{eqnarray}
\chi_3&=& -\frac{1}{3 p_\infty^3}+\frac{4}{p_\infty}+(-8\nu+24) p_\infty\nonumber\\
&+&\left(-36\nu+\frac{64}{3}+8\nu^2\right) p_\infty^3\nonumber\\
&+&\left(-\frac{91}{5}\nu+34\nu^2-8\nu^3\right) p_\infty^5\nonumber\\
&+&\left(\frac{69}{70}\nu+\frac{51}{5}\nu^2-32\nu^3+8\nu^4\right)p_\infty^7 + o(p_\infty^7).
\end{eqnarray}
In this expression the last term $\propto p_\infty^7$ is the 5PN contribution to $\chi_3$. Importantly, we
checked that this newly derived result is in agreement with the corresponding 5PN-level term in the PN expansion
of the (partly conjectural) 3PM-level  recent result of ~\cite{Bern:2019nnu,Bern:2019crd}. This is the first independent, partial confirmation 
of the latter result.

In addition, our results yield an explicit 5PN-accurate value for the 4PM-level scattering angle 
$\chi_4= \chi_4^{\rm loc} + \chi_4^{\rm nonloc}$. Let us only cite here the 5PN-level term in the local contribution 
$\chi_4^{\rm loc}(p_\infty)$:
\begin{eqnarray}
&&{\chi_4}_{\rm loc}^{5 \rm PN}(p_\infty)=\pi \left(-\frac{94899}{32768}\pi^2\nu^2+\frac{93031}{32768}\pi^2\nu\right.\nonumber\\
&&  \quad-\frac{1945583}{33600}\nu+\frac{1937}{16}\nu^2 -\frac{2895}{32}\nu^3+\frac{525}{64}\nu^4\nonumber\\
&& \left.\quad+\frac{1845}{2048}\pi^2\nu^3\right) p_\infty^6 \,.
\end{eqnarray}
The complementary nonlocal contribution is derivable by the methods of ~\cite{Bini:2017wfr}.

 {\it Conclusions.}--- We have introduced a new methodology (based on combining  several different
 theoretical tools) for analytically computing the conservative dynamics of
 two bodies in General Relativity. We have applied our approach to deriving a nearly complete expression for the 5PN-level
 action. It is given by the sum of a 4PN+5PN nonlocal action, Eq. \eqref{Snonloc}, and of a local one $\int pdq- H_{\rm loc}^{\leq {\rm 5PN}} dt$. 
 We determined the full functional structure of $H_{\rm loc}^{\leq {\rm 5PN}}$, except for   two ($\nu^3$-level) unknown coefficients. 
Our results give access to the 5PN-accurate $O(G^3)$ and $O(G^4)$ scattering angles. This provided the
 first independent confirmation of the recent 3PM result of Refs.~\cite{Bern:2019nnu,Bern:2019crd}. 
 
 Our work opens promising avenues for further progress on the dynamics of binary systems. Indeed, the technique we defined here can
 be extended, in principle, to higher PN orders. Our work also offers new motivations for doing targeted, partial computations
 able to determine the two currently missing numerical coefficients. We can think of
 several ways in which they could be determined: second-order self-force computation; partial computation of 5PN dynamics by
 traditional techniques aiming only at terms having selected mass dependence; or, eventually, high-accuracy numerical computation.

\end{document}